\documentclass[amsmath,amssymb]{revtex4}
\usepackage{graphicx}
\usepackage{dcolumn}
\usepackage{bm}
\makeatletter \@addtoreset{equation}{section} \makeatother

\begin{document}
\title{Transversely Stable Soliton Trains in Photonic Lattices}
\author{Jianke Yang }
\affiliation{%
Department of Mathematics and Statistics, University of Vermont,
Burlington, VT 05401
}%

\begin{abstract}
We report the existence of transversely stable soliton trains in
optics. These stable soliton trains are found in two-dimensional
square photonic lattices when they bifurcate from $X$-symmetry
points with saddle-shaped diffraction inside the first Bloch band
and their amplitudes are above a certain threshold. We also show
that soliton trains with low amplitudes or bifurcated from edges of
the first Bloch band ($\Gamma$ and $M$ points) still suffer
transverse instability. These results are obtained in the continuous
lattice model and further corroborated by the discrete model.
\end{abstract}


\maketitle

\section{Introduction}
It is well known that in homogeneous optical media, a bright soliton
stripe, which is localized along the longitudinal direction and
uniform along the transverse (stripe) direction, is always unstable
to transverse perturbations
\cite{Zak_Rub,TI_saturable,Saffman,SHG_TI_neck,SHG_TI_snake,Kiv_Peli,Yang_SIAM}.
When the longitudinal and transverse diffractions have the same
sign, the instability is of neck-type which breaks up the stripe
into filaments \cite{Kiv_Peli,Yang_SIAM,SHG_TI_neck}, while when the
two diffractions have the opposite sign, the instability is of
snake-type which bends the stripe to a wavy form
\cite{Kiv_Peli,Yang_SIAM,SHG_TI_snake}. These two types of
instabilities have been experimentally observed as well
\cite{Saffman,snake}. When a one-dimensional optical lattice is
imposed along the longitudinal or transverse direction of the
soliton stripe, the soliton stripe is still transversely unstable
\cite{Aceves_semi_TI,1Dlattice_TI}. While the transverse instability
can be utilized for certain applications (such as pulse
compression), in many other cases (such as experiments in lower
dimensions), it is detrimental and undesirable. To suppress this
transverse instability, some ideas have been proposed. For instance,
this instability can be completely eliminated if the soliton stripe
is made sufficiently incoherent along the transverse direction
\cite{Segev}. This instability can also be significantly reduced
(but not eliminated) by nonlinearity saturation or incoherent mode
coupling \cite{Segev_Kivshar,Mull_Yang_TI}. Beside optics,
transverse instability is also a common phenomenon in other branches
of physics such as water waves
\cite{Kiv_Peli,KP,Zakharov_KP_TI,waterwave_TI}.

In this paper, we report the existence of coherent and transversely
stable solitons trains in optics (for the first time to our
knowledge). These stable soliton trains are discovered in
two-dimensional photonic lattices. They comprise a periodic sequence
of intensity lumps along the transverse direction, and are localized
along the longitudinal direction. These soliton trains, which are
exact stationary solutions of the underlying nonlinear lattice
system \cite{train_discrete, train_gap, train_ES}, are transversely
stable when they bifurcate from the $X$-symmetry point (with
saddle-shaped diffraction) inside the first Bloch band of the
lattice, and the amplitude of the soliton trains is above a certain
threshold value. This finding is first obtained in the full
continuous lattice model. Then it is also corroborated by the
discrete nonlinear Schr\"odinger model, which is shown to support
transversely stable discrete line solitons as well under similar
conditions. Physically, these soliton trains are transversely stable
due to the stabilizing effect of the photonic-lattice potential.
However, the photonic lattice does not stabilize every soliton
train. Specifically, low-amplitude soliton trains as well as soliton
trains that bifurcate from edges of the first Bloch band ($\Gamma$
and $M$ points) are still transversely unstable. These transversely
unstable soliton trains will also be explained both mathematically
and physically.

\section{Soliton trains in two-dimensional photonic lattices}
\begin{figure}[t]
\includegraphics[width=0.5\textwidth]{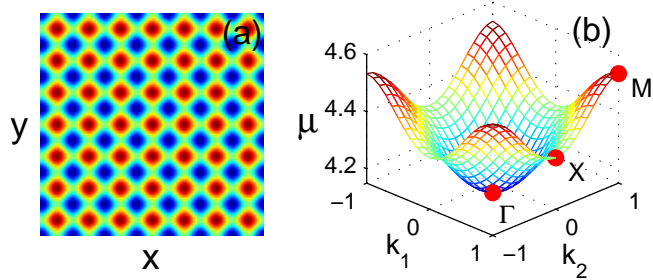}

\caption{(Color online) (a) the photonic lattice (\ref{lattice})
(with $h=6$); (b) the diffraction surface $\mu=\mu(k_1,k_2)$ of the
first Bloch band for this lattice. The high-symmetry points
$\Gamma=(0,0)$, $X=(1,0)$ and $M=(1,1)$ in the irreducible Brillouin
zone $-1\le k_1,k_2\le 1$ are marked by red dots. }
\label{f:lattice}
\end{figure}

The theoretical model we use for coherent-beam propagation in a
two-dimensional (2D) photonic lattice is \cite{Yang_SIAM}
\begin{equation} \label{model}
iU_z+U_{xx}+U_{yy}+n(x,y)U+\sigma |U|^2U=0,
\end{equation}
where $z$ is the direction of propagation, $(x,y)$ is the plane
orthogonal to the propagation direction, $n(x,y)$ is the periodic
refractive-index variation on the orthogonal plane, and $\sigma=\pm
1$ represents self-focusing and self-defocusing nonlinearity. All
variables have been normalized. In our analysis, we take the lattice
to be
\begin{equation} \label{lattice}
n(x,y)=h(\cos^2 \hspace{-0.05cm} x+\cos^2  \hspace{-0.05cm} y),
\end{equation}
where $h$ is the index-variation depth parameter. This is a square
lattice which arises frequently in optics and Bose-Einstein
condensates \cite{Yang_SIAM,BEC_lattice}. This lattice is
$\pi$-periodic along both $x$ and $y$ directions, and is displayed
in Fig. \ref{f:lattice}(a). Below, $x$ and $y$ will be called the
principal axes of the lattice since along them the lattice has the
smallest period. This lattice supports soliton trains aligned along
various directions in the $(x,y)$ plane
\cite{train_discrete,train_gap,train_ES}. In this paper, we only
consider soliton trains alighed along a principal axis of the
lattice for simplicity, and take this principal axis to be the
$y$-direction. Such soliton trains are of the form
\begin{equation}
U(x,y,z)=u(x,y)e^{-i\mu z},
\end{equation}
where $u(x,y)$ is a real-valued function which is localized along
the longitudinal $x$-direction and periodic along the transverse
$y$-direction, and $\mu$ is the propagation constant. The function
$u(x,y)$ satisfies the equation
\begin{equation}  \label{e:u}
u_{xx}+u_{yy}+\mu u+n(x,y)u+\sigma u^3=0.
\end{equation}
In this section, we examine these soliton train solutions.

When $u(x,y)$ is infinitesimal, Eq. (\ref{e:u}) becomes a linear
equation whose bounded solutions are Bloch modes,
\begin{equation} \label{e:Bloch}
p(x,y;\mu)=e^{i(k_1x+k_2y)}\hat{p}(x,y; \mu),
\end{equation}
where $\hat{p}(x,y;\mu)$ is $\pi$-periodic in both $(x,y)$,
$k_1,k_2$ are wavenumbers in the irreducible Brillouin zone $-1\le
k_1,k_2\le 1$, and
\begin{equation}
\mu=\mu(k_1,k_2)
\end{equation}
is the diffraction relation. At the lattice depth $h=6$, this
diffraction function for the first (lowest) Bloch band is shown in
Fig. \ref{f:lattice}(b). If the linear Bloch mode $p(x,y;\mu)$ is
required to be real, this mode then must be at one of the
high-symmetry points, $\Gamma=(0,0)$, $M=(1,1)$, and $X=(1,0),
(0,1)$ in the irreducible Brillouin zone, and be $\pi$- or
$2\pi$-periodic in $x$ and $y$. For the first Bloch band (see Fig.
\ref{f:lattice}(b)), $\Gamma$ and $M$ are the lower and upper edge
points, and $X$ lies inside the Bloch band where the diffraction
surface has a saddle shape. In this paper, we only consider soliton
trains bifurcated from this first band.

When $u(x,y)$ is small but not infinitesimal, Eq. (\ref{e:u}) is
weakly nonlinear, and its solution is a weakly modulated Bloch-wave
packet which bifurcates out from the underlying high-symmetry point
of the Bloch band. When this solution is a soliton train along the
$y$ direction, it can be expanded into a perturbation series
\begin{equation} \label{e:uexpand}
u(x,y)=\epsilon A(X) p(x,y)+\epsilon^2 A'(X)\nu_1(x,y)+\dots,
\end{equation}
\begin{equation}  \label{e:muexpand}
\mu=\mu_0+\tau \epsilon^2,
\end{equation}
where $\mu_0$ is the propagation constant of the high-symmetry
point, $p(x,y)$ is the Bloch wave at $\mu_0$, $\nu_1(x,y)$ is a
generalized Bloch function at $\mu_0$ which satisfies the equation
\begin{equation}  \label{e:nu1}
\partial_{xx}\nu_1+\partial_{yy}\nu_1+[\mu_0+n(x,y)]\nu_1=-2\partial_xp,
\end{equation}
$\tau=\pm 1$, $0<\epsilon\ll 1$, $X=\epsilon x$ is the slow spatial
variable, and $A(X)$ is the one-dimensional (1D) envelope of this
Bloch wave. Notice that the solution $\nu_1(x,y)$ to Eq.
(\ref{e:nu1}) is periodic in both $x$ and $y$ with the same period
as $p(x,y)$. In addition, this solution is not unique since one may
add an arbitrary homogeneous solution $\zeta p(x,y)$, where $\zeta$
is a free constant. Returning to the expansion (\ref{e:uexpand}), we
can see that adding $\zeta p(x,y)$ to $\nu_1(x,y)$ amounts to a
shift in the position of the envelope $A(X)$. In order to fix the
location of the envelope $A(X)$, we require that $\nu_1(x,y)$ be
orthogonal to $p(x,y)$,
\begin{equation}
\int_{-\pi}^{\pi}\int_{-\pi}^{\pi} p(x,y)\nu_1(x,y)dxdy=0.
\end{equation}
This orthogonality requirement uniquely determines the solution
$\nu_1(x,y)$. In the present case, the lattice $n(x,y)$ in
(\ref{lattice}) is symmetric in $x$, then $p(x,y)$ is either
symmetric or antisymmetric in $x$. In this case, $\nu_1(x,y)$ would
have the opposite $x$-symmetry of $p(x,y)$ under the above
orthogonality condition.

By inserting this perturbation expansion into Eq. (\ref{e:u}) and
following the analysis very similar to \cite{Yang_Shi}, we find that
the envelope $A(X)$ satisfies the following equation
\begin{equation} \label{e:A}
D_1A_{XX}+\tau A+\sigma \alpha A^3=0,
\end{equation}
where the $x$-direction diffraction coefficient $D_1$ and the
nonlinear coefficient $\alpha$ are given by
\[
D_1=\left. \frac{1}{2}\frac{\partial^2\mu(k_1,k_2)}{\partial
k_1^2}\right|_{\mu=\mu_0},  \quad
\alpha=\frac{\int_{-\pi}^{\pi}\int_{-\pi}^{\pi}p^4(x,y)
dxdy}{\int_{-\pi}^{\pi}\int_{-\pi}^{\pi}p^2(x,y) dxdy}.
\]
When $\mbox{sgn}(D_1)=\rm{sgn}(\sigma)=-\rm{sgn}(\tau)$, the
envelope equation (\ref{e:A}) admits a sech soliton
\begin{equation}  \label{e:Asech}
A(X)=\sqrt{\frac{2}{|\alpha|}}\hspace{0.1cm}
\mbox{sech}\frac{X-X_0}{\sqrt{|D_1|}},
\end{equation}
where $X_0=\epsilon x_0$ is the location of the peak of the envelope
function $A(X)$. When the Bloch wave $p(x,y)$ is modulated by this
1D sech envelope, the resulting solution (\ref{e:uexpand}) is then a
low-amplitude soliton train along the $y$ direction.

One may notice that the envelope equation (\ref{e:A}) is
translation-invariant, hence $X_0$ is a free parameter in the
envelope solution (\ref{e:Asech}). This seems to imply that soliton
trains can be obtained regardless of the position of the envelope
(\ref{e:Asech}) relative to the underlying periodic potential. This
is not true however. In a similar 1D-lattice model, it has been
shown that the peak of the envelope can only be located at two
positions relative to the lattice \cite{Peli_PRE,Yang_SIAM,Hwang}.
Slight extension of that analysis to the present 2D soliton train
problem shows that the envelope (\ref{e:Asech}) of the soliton
train, for the symmetric cosine lattice (\ref{lattice}), also must
be located at one of the two positions
\begin{equation}
x_0=0, \ \pi/2.
\end{equation}
The resulting soliton train with $x_0=0$ is called the on-site
soliton train, and the other one with $x_0=\pi/2$ is called the
off-site soliton train. The off-site train resides at $(x,y)$
regions of low refractive indices $n(x,y)$ and is thus expected to
be always unstable \cite{Peli_PRE,Yang_SIAM,Hwang}. Thus we only
consider on-site soliton trains in the rest of this article.

\begin{figure}[h]
\includegraphics[width=0.5\textwidth]{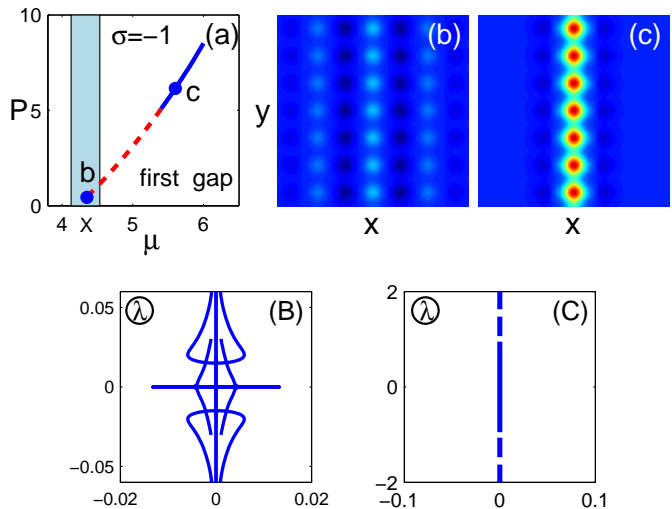}

\caption{(Color online) (a) the power curve of soliton trains
bifurcated from the $X$ point inside the first Bloch band under
defocusing nonlinearity; the dashed (red) segment is transversely
unstable, while the solid (blue) segment is stable; (b,c) profiles
$u(x,y)$ of soliton trains at low and high amplitudes respectively;
these solutions are located at the points marked by the same letters
on the power curve of (a); (B,C) linear-stability spectra for the
soliton trains in (b,c). } \label{f:Xfamily}
\end{figure}

When the amplitude of the soliton train is not small, the above
perturbation series would be invalid. In such cases, soliton trains
can be determined numerically \cite{Yang_SIAM}. To illustrate, we
take self-defocusing nonlinearity ($\sigma=-1$) and lattice depth
$h=6$. In this case, a family of soliton trains bifurcates out from
the $X$ point inside the first Bloch band. The power curve of this
family is displayed in Fig. \ref{f:Xfamily}(a). Here the power $P$
is defined as the integral of $u^2$ from $-\infty<x<\infty$ and on
one transverse period $0\le y\le\pi$, i.e.,
\begin{equation}
P=\int_{-\infty}^\infty dx \int_{0}^\pi dy \ u^2(x,y).
\end{equation}
Two typical soliton trains, with low and high amplitudes (powers),
are displayed in Fig. \ref{f:Xfamily}(b,c) respectively. The
propagation constants for these two soliton trains are $\mu=4.3495$
and 5.6. One can see that the intensity peaks of these solitons are
in-phase along the transverse ($y$) direction and out-phase along
the longitudinal ($x$) direction. In addition, the soliton train
near the $X$ point has low amplitude and power, and is
longitudinally broad (occupying many lattice sites), while that away
from the $X$ point has high amplitude and power, and is
longitudinally strongly localized (occupying practically a single
lattice site). This type of soliton trains has been theoretically
predicted and experimentally observed in \cite{train_ES}. In Fig.
\ref{f:Mfamily} of later text, another family of soliton trains
bifurcated from the $M$-symmetry point of the first Bloch band will
also be displayed.

In the next section, we will examine the transverse stability of
soliton trains. Before detailed analysis, let us first develop some
intuition. The soliton trains exist under both self-focusing and
self-defocusing nonlinearities ($\sigma=\pm 1$), and they can
bifurcate from the $\Gamma$ and $M$ points at edges of the Bloch
band \cite{train_discrete,train_gap}, or from $X$ points inside the
Bloch band \cite{train_ES}. For soliton trains bifurcated from the
$\Gamma$ point (which occurs under self-focusing nonlinearity)
\cite{train_discrete}, the intensity peaks along the transverse
(train) direction are all in-phase. For soliton trains bifurcated
from the $M$ point (which occurs under self-defocusing nonlinearity)
\cite{train_gap}, the adjacent intensity peaks along the transverse
direction are all out of phase (if this transverse direction is
along a principal axis of the lattice as in our present case, see
Fig. \ref{f:Mfamily}). It is known that in-phase dipoles under
self-focusing nonlinearity and out-phase dipoles along a principal
axis of the lattice under self-defocusing nonlinearity are both
unstable
\cite{Yang_dipole,Chen_dipole,Kivshar_transform,Yang_SIAM,Kevrekidis_book}.
Then if the soliton trains are strongly localized along the
longitudinal direction (which occurs at large amplitudes, see Fig.
\ref{f:Mfamily}(c)), we may view the soliton trains as a collection
of transverse dipoles, hence we may expect the soliton trains
bifurcated from $\Gamma$ and $M$ points to be unstable. However, for
soliton trains bifurcated from the $X$ points (under either
self-focusing or self-defocusing nonlinearity), adjacent transverse
intensity peaks are out of phase under self-focusing nonlinearity,
and are in-phase under self-defocusing nonlinearity (see Fig.
\ref{f:Xfamily} and \cite{train_ES}). Dipoles with such phase
structures are stable in deep lattices
\cite{Yang_dipole,Chen_dipole,Kivshar_transform,Yang_SIAM,Kevrekidis_book}.
Thus soliton trains bifurcated from $X$ points (at high amplitudes
as in Fig. \ref{f:Xfamily}(c)) may be free of transverse
instabilities. In the next two sections, we will confirm that these
intuitions are largely correct, hence transversely stable soliton
trains will be discovered.

It should be cautioned, however, that these intuitions are
reasonable only when the soliton trains have high amplitudes, so
that they are strongly longitudinally localized (as in Figs.
\ref{f:Xfamily}(c) and \ref{f:Mfamily}(c)), which makes the
transverse-dipole analogy meaningful. If the soliton trains have low
amplitude, then they are longitudinally broad and occupying many
lattice sites (see Fig. \ref{f:Xfamily}(b)). In that case,
longitudinal inter-site coupling becomes important, which makes the
above transverse-dipole analogy inappropriate. As we will show
later, all low-amplitude soliton trains are transversely unstable.

\section{Transversely stable soliton trains}

In this section, we study the transverse stability of soliton trains
in 2D photonic lattices, and show that transversely stable soliton
trains exist.

First, we briefly show that \emph{low-amplitude} soliton trains are
always transversely unstable. To show this, we consider the dynamics
of a low-amplitude soliton train, whose solution can be expanded
into a perturbation series,
\begin{equation} \label{e:Uexpand}
U(x,y,z)=e^{-i\mu_0z}\left[\epsilon \Psi(X,Y,Z) p(x,y)+\epsilon^2
U_2 + \dots\right],
\end{equation}
where $\mu_0$ is the propagation constant of a high-symmetry point,
$p(x,y)$ is the Bloch wave at $\mu_0$, $0<\epsilon\ll 1$,
$X=\epsilon x$, $Y=\epsilon y$ are slow spatial variables,
$Z=\epsilon^2z$ is the slow propagation-distance variable, and
$\Psi(X,Y,Z)$ is the envelope function of this low-amplitude soliton
train. Following the analysis of \cite{Yang_Shi}, it is easy to show
that the evolution of the envelope function $\Psi(X,Y,Z)$ is
governed by the following constant-coefficient 2D NLS equation
\begin{equation}  \label{e:Psi}
i\Psi_Z+D_1\Psi_{XX}+D_2\Psi_{YY}+\sigma\alpha |\Psi|^2\Psi=0,
\end{equation}
where $D_2$ is the $y$-direction diffraction coefficient,
\[
D_2=\left. \frac{1}{2}\frac{\partial^2\mu(k_1,k_2)}{\partial
k_2^2}\right|_{\mu=\mu_0},
\]
and $D_1, \alpha$ have been given before. This 2D envelope equation
admits a line-soliton solution
\begin{equation} \label{e:line_soliton}
\Psi(X,Y,Z)=A(X)e^{-i\tau Z},
\end{equation}
where $\tau=-\sigma$, and $A(X)$ is given in Eq. (\ref{e:Asech}).
This envelope line soliton, when modulated onto the Bloch mode
$p(x,y)$, yields the low-amplitude soliton train (\ref{e:uexpand})
derived in the previous section. It is well known that this envelope
line soliton (\ref{e:line_soliton}) is transversely unstable in the
constant-coefficient 2D envelope equation (\ref{e:Psi})
\cite{Zak_Rub,Kiv_Peli,Yang_SIAM}. Thus low-amplitude soliton trains
(\ref{e:uexpand}) are also transversely unstable. Since the
transverse instability of low-amplitude soliton trains is induced by
the transverse instability of their envelope line solitons,
obviously the longitudinal wave coupling plays an important role in
this instability (as has been pointed out in the end of the previous
section).

We can further derive the analytical formula for unstable
eigenvalues of low-amplitude soliton trains. Since the soliton train
is periodic along the transverse direction, the normal modes of
infinitesimal disturbances to the train are of the form
$\widetilde{\Psi} \sim e^{iky+\lambda z}\psi(x,y)$, where $k$ is the
transverse wavenumber of the disturbance, $\lambda$ is the
eigenvalue, and $\psi(x,y)$ is $\pi$-periodic in $y$. From the above
analysis and after variable scalings, it is easy to show that the
eigenvalue $\lambda(k)$ is given by
\begin{equation} \label{e:formulalambda}
\lambda(k)=\epsilon^2\Lambda(\sqrt{|D_2|} \hspace{0.08cm}
k/\epsilon),
\end{equation}
where $\Lambda(K)$ is the eigenvalue of the normalized sech line
soliton
\begin{equation}
\Phi_0(X,Y,Z)=\sqrt{2}\ \mbox{sech}X \hspace{0.03cm} e^{iZ}
\end{equation}
in the \emph{unit-coefficient} 2D NLS equation
\begin{equation}  \label{e:Phi}
i\Phi_Z+\Phi_{XX}+\mbox{sgn}(D_1D_2) \Phi_{YY}+ |\Phi|^2\Phi=0
\end{equation}
for the disturbance proportional to $e^{iKY}$, which have been
obtained in \cite{Yang_SIAM,Deconnick}. A more rigorous derivation
of the eigenvalue formula (\ref{e:formulalambda}), which is based on
the study of the linear-stability eigenvalue problem of
low-amplitude soliton trains, can also be made by a slight
modification of the analysis in \cite{Shi_stability}.

If the low-amplitude soliton train bifurcates from the $X$-symmetry
point (as in Fig. {\ref{f:Xfamily}), $\mbox{sgn}(D_1D_2)=-1$, hence
the envelope's transverse instability is of snake type. Since
$D_1D_2<0$, the diffraction surface at the $X$ point has a saddle
shape (see Fig. \ref{f:lattice}(b)). If the train bifurcates from
the edges of the first Bloch band (as in Fig. \ref{f:Mfamily}),
$\mbox{sgn}(D_1D_2)=1$, hence the envelope's transverse instability
is of neck type.

When the amplitude of the soliton train is not small, the analytical
stability analysis above becomes invalid. Below, we numerically
determine the transverse stability of higher-amplitude soliton
trains (using numerical methods described in \cite{Yang_SIAM}).

First, we consider the soliton trains in Fig. \ref{f:Xfamily}, which
bifurcate from the $X$-symmetry point of the first Bloch band. The
numerically obtained linear-stability spectra for the low- and
high-amplitude soliton trains of Fig. \ref{f:Xfamily}(b,c) are
displayed in Fig. \ref{f:Xfamily}(B,C) respectively. The spectrum in
Fig. \ref{f:Xfamily}(B) contains both real and complex unstable
eigenvalues which lie on the right half of the spectral
$\lambda$-plane, indicating that the low-amplitude soliton train in
Fig. \ref{f:Xfamily}(b) is transversely unstable. This numerical
result agrees with the analytical result given above. Quantitatively
we have also compared these numerical eigenvalues with the
analytical formula (\ref{e:formulalambda}), with the function
$\Lambda(K)$ provided in \cite{Yang_SIAM,Deconnick}, and found
excellent quantitative agreement as well.

A more important finding on this family of soliton trains is that,
when the amplitude (or power) of these soliton trains reaches above
a certain threshold, transverse instability disappears, and these
soliton trains become fully stable. This can be seen in Fig.
\ref{f:Xfamily}(C), which gives the linear-stability spectrum for
the high-amplitude soliton train in Fig. \ref{f:Xfamily}(c). This
spectrum does not contain any unstable eigenvalue, indicating that
this high-amplitude soliton train is fully stable. What happens is
that, as $\mu$ moves away from the $X$ point $\mu_0=4.330$ (i.e., as
$\epsilon$ increases), the unstable eigenvalues initially increase
in size and drift away from the imaginary axis of the spectral
plane, as is predicted by the analytical formula
(\ref{e:formulalambda}). However, as $\mu$ increases further, the
unstable eigenvalues turn around and start to move toward the
imaginary axis (hence the transverse instability weakens). When
$\mu>\mu_c\approx 5.44$, or $P>5.3$, all unstable eigenvalues merge
into the imaginary axis, hence instability vanishes, and the soliton
trains become transversely stable. Physically, what happens is that,
as $\mu$ moves away from the $X$ point, the soliton train transforms
from a low-amplitude longitudinally-broad train into a
high-amplitude longitudinally-narrow train (see Fig.
\ref{f:Xfamily}(b,c)). In this process, the snaking instability of
the low-amplitude longitudinally-broad soliton train is eventually
arrested by the lattice when the train becomes strongly localized
longitudinally.

We have found that the high-amplitude soliton train in Fig.
\ref{f:Xfamily}(c) is not only linearly stable, but also nonlinearly
stable. To demonstrate its nonlinear stability, we simulate its
evolution in Eq. (\ref{model}) when it is perturbed by 10\%
random-noise transverse perturbations. After long-distance
simulations, we have found that this train remains robust and does
not break up at all. For instance, one simulation result (with
simulation distance $z=100$) is displayed in Fig. \ref{f:evolution}.

\begin{figure}
\includegraphics[width=0.3\textwidth]{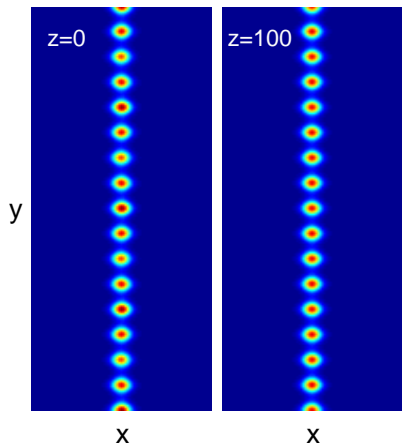}

\caption{(Color online) Nonlinear evolution of the soliton train in
Fig. \ref{f:Xfamily}(c) under 10\% initial transverse perturbations
(shown are intensity fields $|U|^2$). Left: the initial perturbed
soliton train; right: the solution at $z=100$. } \label{f:evolution}
\end{figure}

The soliton train in Fig. \ref{f:Xfamily}(c) may remind us the train
of lumps which form after the onset of neck-type transverse
instability to a line soliton in a homogeneous medium
\cite{Saffman,Yang_SIAM}. However, it is important to recognize that
the train of lumps in a homogeneous medium is not a stable
structure. Upon further propagation, it will break up
\cite{Yang_SIAM}. In contrast, the soliton train in Fig.
\ref{f:Xfamily}(c) is both linearly and nonlinearly stable, and can
propagate for all distances without breakup (see Fig.
\ref{f:evolution}). Here the photonic lattice plays an important
role in the stabilization of the soliton train in the presence of
transverse perturbations.

The photonic lattice, however, cannot stabilize every soliton train
(even if their amplitude is high). To demonstrate, we consider the
soliton trains which bifurcate from the upper edge of the first
Bloch band ($M$-symmetry point) under self-defocusing nonlinearity
(with $h=6$ as before). This soliton family is displayed in Fig.
\ref{f:Mfamily}(a,c). From the solution profile shown in Fig.
\ref{f:Mfamily}(c), we see that the intensity peaks of these soliton
trains are out of phase with each other along both the transverse
and longitudinal directions (as is expected since these trains
bifurcate from the $M$-symmetry point). However, this family of
soliton trains are all transversely unstable. To demonstrate, the
linear-stability spectrum for the high-amplitude soliton train in
Fig. \ref{f:Mfamily}(c) is shown in Fig. \ref{f:Mfamily}(b). The
real unstable eigenvalues on the right half of the spectral plane
indicate that this high-amplitude soliton train is linearly
unstable. The nonlinear instability of this soliton train is
displayed in Fig. \ref{f:Mfamily}(d). It is seen that under weak
perturbations, this soliton train breaks up into filaments. For this
solution family, the transverse instability of low-amplitude soliton
trains is neck-type (since the diffraction coefficients $D_1,D_2$
have the same sign at the $M$ point). Then Fig. \ref{f:Mfamily}
shows that the neck-type transverse instability of low-amplitude
soliton trains cannot be arrested by the photonic lattice as the
soliton train's amplitude becomes high.

From Figs. \ref{f:Xfamily} and  \ref{f:Mfamily}, we learn that the
phase relation of transverse intensity humps also plays an important
role in the stabilization of soliton trains. Specifically, under
defocusing/focusing nonlinearity, the transverse intensity peaks of
the soliton train should be in-phase/out-phase in order for it to be
stable (as one would expect from dipole stability results
\cite{Yang_dipole,Chen_dipole,Kivshar_transform,Yang_SIAM,Kevrekidis_book}).
This requirement on the phase relation translates into a requirement
that these soliton trains should bifurcate from the $X$-symmetry
point inside the Bloch band.

\begin{figure}
\includegraphics[width=0.48\textwidth]{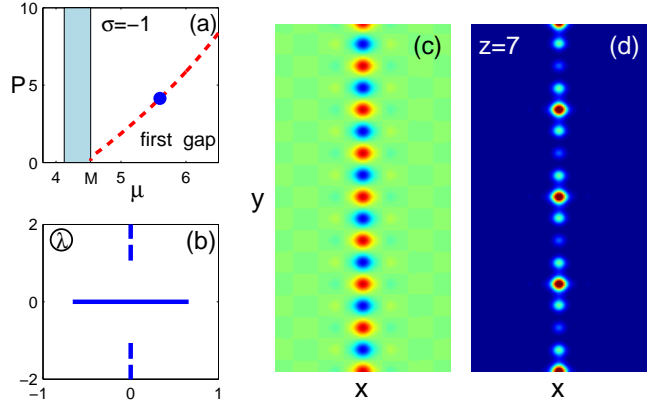}

\caption{(Color online) (a) the power curve of the soliton trains
bifurcated from the $M$ point of the first Bloch band under
self-defocusing nonlinearity (the dashed red line indicates this
whole family is transversely unstable); (b,c) the stability-spectrum
and profile $u(x,y)$ of the soliton train at the marked point on the
power curve ($\mu=5.6$); (d) nonlinear evolution ($z=7$) of the
soliton train in (c) under 10\% initial transverse perturbations
(shown is the intensity field $|U|^2$). } \label{f:Mfamily}
\end{figure}

We should add that, for the family of soliton trains in Fig.
\ref{f:Xfamily}, even though they become stable when
$\mu>\mu_c\approx 5.44$, if $\mu$ gets close to the second Bloch
band (whose lower edge is at $\mu=7.23$), then these soliton trains
would become unstable again due to wave coupling to the second band.

\section{Transversely stable line solitons in the discrete NLS
equation}

In the previous section, we demonstrated the existence of
transversely stable soliton trains in the continuous lattice model
(\ref{model}). In this section, we corroborate this finding by
showing that the analogous transversely stable line solitons exist
in the discrete NLS model as well. The discrete NLS model is often
used to qualitatively describe wave dynamics in the continuous
lattice model (\ref{model}) with a deep lattice potential under the
tight-binding approximation \cite{Kevrekidis_book}. In this context,
the variable in the discrete model can be viewed as the complex
amplitude of the ground-state eigenmode of each lattice-cell
potential. The approximation of the continuous model (\ref{model})
by the discrete NLS equation is a significant reduction. Under this
reduction, the soliton train shown in Fig. \ref{f:Xfamily}(b,c) in
the continuous model becomes a transversely \emph{uniform} line
soliton in the discrete model (see Fig. \ref{f:Xfamilyd}(c,d)
below). This treatment simplifies both analytical analysis and
numerical computations.

The discrete NLS equation we consider is
\begin{equation} \label{e:DNLS}
i\frac{d}{dz}U_{m,n}+\Delta_2U_{m,n}+\sigma |U_{m,n}|^2U_{m,n}=0,
\end{equation}
where $\Delta_2$ is the two-dimensional difference operator
\begin{eqnarray}
\Delta_2U_{m,n} & \equiv & (U_{m+1,n}-2U_{m,n}+U_{m-1,n})+  \nonumber \\
&& (U_{m,n+1}-2U_{m,n}+U_{m,n-1}),  \nonumber
\end{eqnarray}
and $\sigma=\pm 1$ is the sign of nonlinearity. Here the
intersite-coupling coefficient in front of $\Delta_2U_{m,n}$ has
been normalized to be unity through scalings of $z$ and $U_{m,n}$.
An important property of this discrete model is that self-focusing
and self-defocusing nonlinearities can be transformed to each other
since this model is invariant under the transformation
\cite{Kivshar_transform}
\begin{equation}
U_{m,n} \to (-1)^{m+n}e^{-8iz}U^*_{m,n}, \quad \sigma \to -\sigma.
\end{equation}
This transformation is very helpful for us to understand the
connection on soliton configurations and their stability properties
between self-focusing and self-defocusing nonlinearities. It also
means that one only needs to study one type of nonlinearity (say,
self-defocusing nonlinearity), and infer the results for the other
type of nonlinearity by this transformation.

Solitons in the discrete model (\ref{e:DNLS}) are sought in the form
\begin{equation}
U_{m,n}(z)=u_{m,n}e^{-i\mu z},
\end{equation}
where $u_{m,n}$ is a real-valued function which satisfies the
equation
\begin{equation} \label{e:umn}
\Delta_2u_{m,n}+\mu u_{m,n}+\sigma u_{m,n}^3=0,
\end{equation}
and $\mu$ is the propagation constant. When $u_{m,n}$ is
infinitesimal, the nonlinear term in (\ref{e:umn}) drops out, and
the bounded solution to the remaining linear equation is then a
discrete Fourier mode, $u_{m,n}=e^{i(k_1m+k_2n)}$, where $-\pi \le
k_1, k_2\le \pi$ are wavenumbers along the $m$- and $n$-directions.
Inserting this discrete Fourier mode into the linear part of Eq.
(\ref{e:umn}), we get the diffraction relation
\begin{equation} \label{e:discrete_diffraction}
\mu=2(2-\cos k_1-\cos k_2).
\end{equation}
The corresponding diffraction surface is shown in Fig.
\ref{f:Xfamilyd}(a). It is easy to see that this diffraction surface
closely resembles the first Bloch band of the continuous model (see
Fig. \ref{f:lattice}). Thus the discrete model (\ref{e:DNLS}) is
appropriate for describing wave dynamics associated with the first
Bloch band in the continuous model (\ref{model}). Notice that the
continuous spectrum of this discrete model (\ref{e:DNLS}) contains
only a single band $0\le\mu \le 8$, while the continuous spectrum of
the continuous model (\ref{model}) often contains multiple Bloch
bands. Thus if wave dynamics in the continuous model (\ref{model})
involves higher Bloch bands, the discrete model (\ref{e:DNLS}) will
be inappropriate.

\begin{figure}
\includegraphics[width=0.48\textwidth]{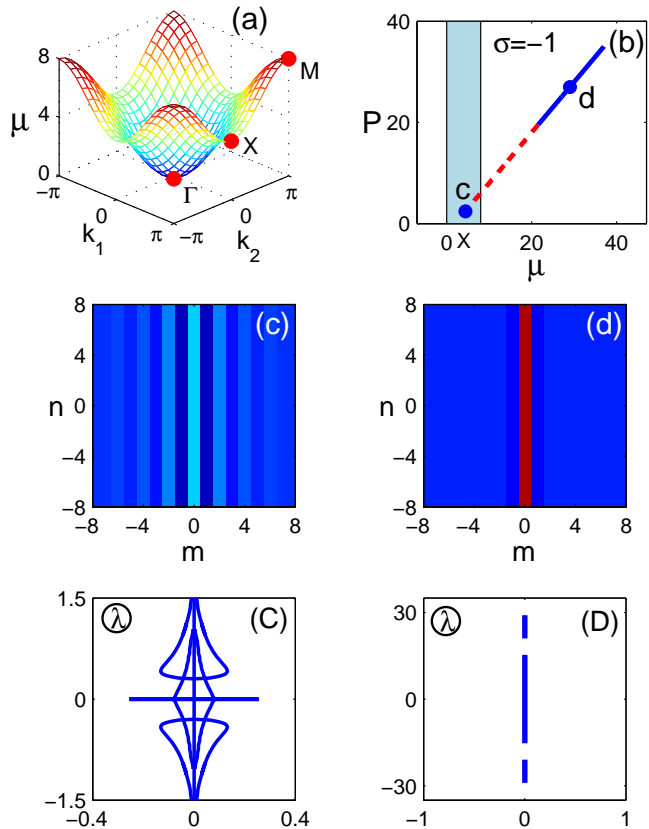}

\caption{(Color online) (a) the diffraction surface
(\ref{e:discrete_diffraction}) of the discrete NLS equation; (b) the
power curve of discrete line solitons bifurcated from the $X$ point
inside the continuum band under self-defocusing nonlinearity; the
dashed (red) segment is transversely unstable, while the solid
(blue) segment is stable; (c,d) profiles $u_{m,n}$ of discrete line
solitons at low and high amplitudes respectively; these solitons are
located at the points marked by the same letters on the power curve
of (b); (C,D) linear-stability spectra for the discrete line
solitons in (c,d).  } \label{f:Xfamilyd}
\end{figure}

Corresponding to the continuous soliton trains in Fig.
\ref{f:Xfamily}, we can find a family of discrete line solitons that
bifurcate from the $X$-symmetry point of the continuum band under
self-defocusing nonlinearity ($\sigma=-1$). These discrete line
solitons $u_{m,n}$ are $n$-independent, thus they are simply 1D
discrete solitons in the 2D model (\ref{e:umn}). The power curve of
this soliton family is shown in Fig. \ref{f:Xfamilyd}(b). Here the
power $P$ is defined as
\begin{equation}
P(\mu)=\sum_{m=-\infty}^\infty u^2_{m,n}.
\end{equation}
Profiles of two typical solitons, near and far away from the $X$
point (with $\mu=4.4$ and 29), are displayed in Fig.
\ref{f:Xfamilyd}(c,d). The one in Fig. \ref{f:Xfamilyd}(c) has low
amplitude (power) and is longitudinally broad, which is the
counterpart of the low-amplitude soliton train in Fig.
\ref{f:Xfamily}(b). The one in Fig. \ref{f:Xfamilyd}(d) has high
amplitude (power) and is longitudinally strongly localized, which is
the counterpart of the high-amplitude soliton train in Fig.
\ref{f:Xfamily}(c). To determine the transverse stability of these
discrete line solitons, we have computed their linear-stability
spectra, and the results are displayed in Fig.
\ref{f:Xfamilyd}(C,D). The spectrum in Fig. \ref{f:Xfamilyd}(C)
indicates that the low-amplitude discrete line soliton in Fig.
\ref{f:Xfamilyd}(c) is transversely unstable. In addition, this
spectrum qualitatively closely resembles the continuous counterpart
in Fig. \ref{f:Xfamily}(B). The spectrum in Fig.
\ref{f:Xfamilyd}(D), on the other hand, does not contain any
unstable eigenvalue, indicating that the high-amplitude discrete
line soliton in Fig. \ref{f:Xfamilyd}(d) is transversely stable.
Thus the existence of transversely stable discrete line solitons is
established. In the present discrete model (\ref{e:DNLS}), the
threshold for transversely stable line solitons is at
$\mu>\mu_c\approx 21.5$, or $P>39.0$.

Now we examine nonlinear developments of these discrete line
solitons under transverse perturbations. We find that when these
solitons are linearly unstable (see Fig. \ref{f:Xfamilyd}(b)), then
under perturbations, they would develop snake instability and
eventually disperse away. This is illustrated in Fig.
\ref{f:evolutiond}(a,b) for the low-amplitude discrete line soliton
in Fig. \ref{f:Xfamilyd}(c). Analytically, this snake instability
can also be understood. Briefly speaking, the envelope of a
low-amplitude discrete solution to Eq. (\ref{e:DNLS}) is governed by
an equation similar to (\ref{e:Psi}), where $D_1$ and $D_2$ are the
diffraction coefficients. At the $X$-symmetry point, $D_1D_2<0$,
thus the line soliton to this envelope equation suffers snake-type
instability \cite{Zak_Rub,Kiv_Peli,Yang_SIAM}, which translates to
the snake instability observed in Fig. \ref{f:evolutiond}(a,b). When
the discrete line solitons are linearly stable (see Fig.
\ref{f:Xfamilyd}(b)), however, they would propagate robustly for all
distances without breakup. This is illustrated in Fig.
\ref{f:evolutiond}(c,d) for the high-amplitude discrete line soliton
in Fig. \ref{f:Xfamilyd}(d).

\begin{figure}[h]
\includegraphics[width=0.35\textwidth]{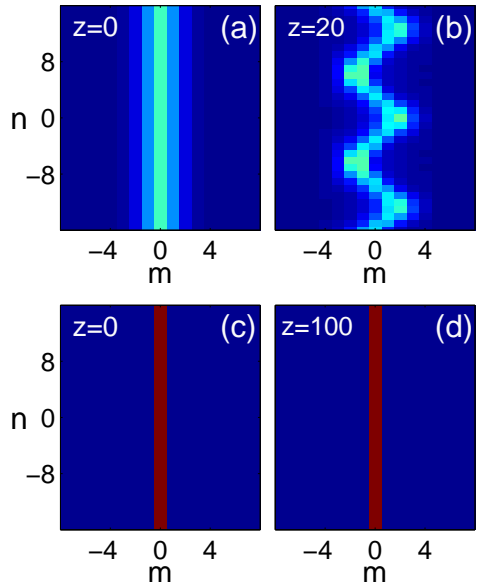}

\caption{(Color online) Nonlinear evolutions of discrete line
solitons in Fig. \ref{f:Xfamilyd}(c,d) under 10\% initial transverse
perturbations (shown are intensity fields $|U|^2$). Upper row: the
low-amplitude case; lower row: the high-amplitude case.}
\label{f:evolutiond}
\end{figure}

By comparing these results for the discrete NLS model (\ref{e:DNLS})
to those for the continuous model (\ref{model}), one can easily see
that the results for both models are qualitatively almost identical.
For both models, we discovered transversely stable soliton trains
under similar conditions, i.e., when they bifurcate from
$X$-symmetry points and have high amplitudes. The instability
behaviors for low-amplitude soliton trains are also the same in both
models. However, minor differences between the two models do exist.
For instance, in the discrete model, line solitons bifurcated from
the $X$ point in Fig. \ref{f:Xfamilyd} are transversely stable for
all $\mu>\mu_c\approx 21.5$; while in the continuous model, soliton
trains bifurcated from the $X$ point in Fig. \ref{f:Xfamily}(a) can
become unstable again when $\mu$ gets close to the second Bloch
band. The reason for this difference is that the discrete model can
not capture the multi-band coupling, as we have mentioned earlier in
this section.

\section{Summary}

\vspace{0.15cm} In summary, we have reported the existence of
transversely stable soliton trains in optics. These soliton trains
are found in two-dimensional square photonic lattices when they
bifurcate from $X$ points (with saddle-shaped diffraction) inside
the first Bloch band, and their amplitudes are above a certain
threshold. These stable soliton trains arise due to the combined
effect of the photonic lattice, proper transverse phase relation,
and strong longitudinal localization. We have also shown that
soliton trains with low amplitudes or bifurcated from edges of the
first Bloch band ($\Gamma$ and $M$ points) still suffer transverse
instability. These results have been obtained in both the continuous
lattice model and the discrete NLS model, and results from both
models are in very good qualitative agreement.

\section*{Acknowledgment} This work is supported in part by the Air Force
Office of Scientific Research (Grant USAF 9550-09-1-0228) and the
National Science Foundation (Grant DMS-0908167).

\end{document}